\documentclass[aps,preprint,pre,superscriptaddress,runinaddress,showpacs]{revtex4}

\usepackage{amssymb}
\usepackage{amsmath}
\usepackage{graphicx}
\usepackage{bm}

\begin{document}

\title{Shell Structure of Confined Charges at Strong Coupling}
\author{J. Wrighton}
\author{J. Dufty}
\affiliation{Department of Physics, University of Florida, Gainesville, FL 32611}
\author{M. Bonitz}
\author{H. K\"ahlert}
\affiliation{Institut f\"{u}r Theoretische Physik und Astrophysik, Christian-Albrechts
Universit\"{a}t zu Kiel, 24098 Kiel, Germany}
\date{\today}

\begin{abstract}
A theoretical description of shell structure for charged particles in a
harmonic trap is explored at strong coupling conditions of $\Gamma =50$ and $%
100$. The theory is based on an extension of the hypernetted chain
approximation to confined systems plus a phenomenological representation of
associated bridge functions. Predictions are compared to corresponding Monte
Carlo simulations and quantitative agreement for the radial density profile
is obtained.
\end{abstract}

\maketitle

Systems of harmonically trapped charged particles exhibit a shell structure
in their radial density profile. Recent studies \cite%
{Pohl2004,Arp,Ludwig2005,bonitz2006,Golubnychiy2006,Baumgartner},
both experimental and molecular dynamics simulation, indicate the
localization of the charges on the surfaces of concentric spheres
with a crystal-like ordering on each surface depending on the
particle number $N$. These are effectively the zero temperature
ground state for the system or energetically close excited states.
An accurate and instructive theoretical description of these
spherical crystals has been given recently using shell models
\cite{Baumgartner,Kraeft06,Cioslowski08}. A complementary
theoretical description of the fluid phase at finite temperatures
has now been given as well \cite{Wrighton09}, showing a closely
related "blurred" shell structure that sharpens at stronger Coulomb
coupling $\Gamma \equiv q^{2}/r_{0}k_{B}T$ ($q$ is the charge,
$r_{0} $ is the ion sphere radius in terms of the average trap
density). This fluid phase theory demonstrates that correlations
play the dominant role in the formation of shell structure. An
extension of the hypernetted chain approximation (HNC) for bulk
fluids to the localized charges of a trap was shown to provide all
of the qualitative features of shell structure (e.g., number and
location of density peaks, shell populations) for the Coulomb
coupling constant $\Gamma \geq 10$ . However, comparison with Monte
Carlo (MC) simulation results at selected conditions showed
significant errors in the HNC amplitudes and widths of the shells
($20-40\%$). An adjusted HNC (AHNC) was proposed to correct these
deficiencies, using a phenomenological representation of the
neglected "bridge functions". Excellent agreement with MC results
was obtained in this way for $\Gamma =20,N=25,100$ and $\Gamma
=40,N=300$. The objective here is to
report further exploration of the AHNC at the stronger coupling values $%
\Gamma =50$ and $100$, and to note some interesting similarities to the
crystal shell structure. \

The origin of the AHNC theory for confined systems is summarized briefly
first. Consider a system of $N$ classical charges in an external potential.
The Hamiltonian for this system is

\begin{equation}
H=H_{0}+\sum_{i=1}^{N}V_{0}(\mathbf{r}_{i}),\hspace{0.25in}%
H_{0}=\sum_{i=1}^{N}\frac{1}{2}mv_{i}^{2}+\frac{1}{2}\sum_{i\neq
j=1}^{N}V(r_{ij})  \label{1}
\end{equation}%
where $\vec{r}_{i}$ and $\vec{v}_{i}$ are the position and velocity of
charge $i$. The external potential seen by each \ particle is denoted by $%
V_{0}(r)$, and the interaction between the pair $i,j$ is $V(r_{ij})$
(application here will be limited to Coulomb interactions but the
discussion is more general). The external potential induces a
non-uniform equilibrium
density $n(\mathbf{r})$. It follows from density functional theory that $n(%
\mathbf{r})$ obeys the equation \cite{Evans}%
\begin{equation}
\ln \frac{n(\mathbf{r})\lambda ^{3}}{z}=-\beta V_{0}\left( \mathbf{r}\right)
-\beta \frac{\delta F_{ex}\left( \beta \mid n\right) }{\delta n\left(
r\right) },  \label{2}
\end{equation}%
where $z=e^{\beta \mu }$, $\mu $ is the chemical potential, and $\lambda
=\left( h^{2}\beta /2\pi m\right) ^{1/2}$ is the thermal wavelength. The
excess free energy $F_{ex}\left( \beta \mid n\right) $ is a universal
functional of the density for the Hamiltonian $H_{0}$, independent of the
applied external potential $V_{0}$, and describes all correlations among the
particles. The solutions to (\ref{2}) are such that there is a unique
equilibrium density $n(\mathbf{r})$ for each $V_{0}(\mathbf{r})$, using the
same $F_{ex}\left( \beta \mid n\right) $. A special case is the uniform
density of a one component plasma (OCP), resulting from the $V_{0}\left(
\mathbf{r}\right) $ for a uniform neutralizing background. The quite
different non-uniform density of interest here results from the harmonic
trap $V_{0}\left( \mathbf{r}\right) =m\omega ^{2}r^{2}/2$. (The spherical
symmetry of $V_{0}$ in both cases yields a spherically symmetric density
profile in the fluid phase; the broken symmetry crystal profile is not
considered here).

This observation that the OCP and trap densities are determined from the
same excess free energy functional opens the possibility of describing
correlations for the trap in terms of those of the OCP. This is done in
reference \cite{Wrighton09} with the result

\begin{equation}
\ln \frac{n(r)\lambda ^{3}}{z}=-\beta \frac{1}{2}m\omega ^{2}r^{2}+\beta
\int d\vec{r}^{\prime }\ n(r^{\prime })c_{\mathrm{OCP}}\left( \vec{r}-\vec{r}%
^{\prime }|;\Gamma \right) +B_{\mathrm{T}}(r|n)  \label{3}
\end{equation}%
where $c_{\mathrm{OCP}}$ is the direct correlation function of a
one-component plasma, and $B_{\mathrm{T}}(r|n)$ is the \textquotedblleft
bridge function" for the trapping potential. The Ornstein-Zernicke equation
relates $c_{\mathrm{OCP}}$ to the pair distribution function for the OCP, $%
g_{\mathrm{OCP}}$. A similar analysis gives an equation for $g_{\mathrm{OCP}}
$

\begin{equation}
\ln g_{\mathrm{OCP}}(r)=-\beta q^{2}r^{-1}+\beta \int d\vec{r}^{\prime }\ %
\left[ g_{\mathrm{OCP}}(r^{\prime })-1\right] c_{\mathrm{OCP}}\left( |\vec{r}%
-\vec{r}^{\prime }|;\Gamma \right) +B_{\mathrm{OCP}}(r|g_{\mathrm{OCP}}).
\label{4}
\end{equation}%
These two equations provide a formally exact description of the charged
particle system from which approximations can be made. In particular, the
HNC approximation for both the OCP and the trap density is defined by the
neglect of both bridge functions in these equations. The trap density is
then determined entirely in terms of correlations for the OCP.

\begin{figure}[tbp]
\includegraphics{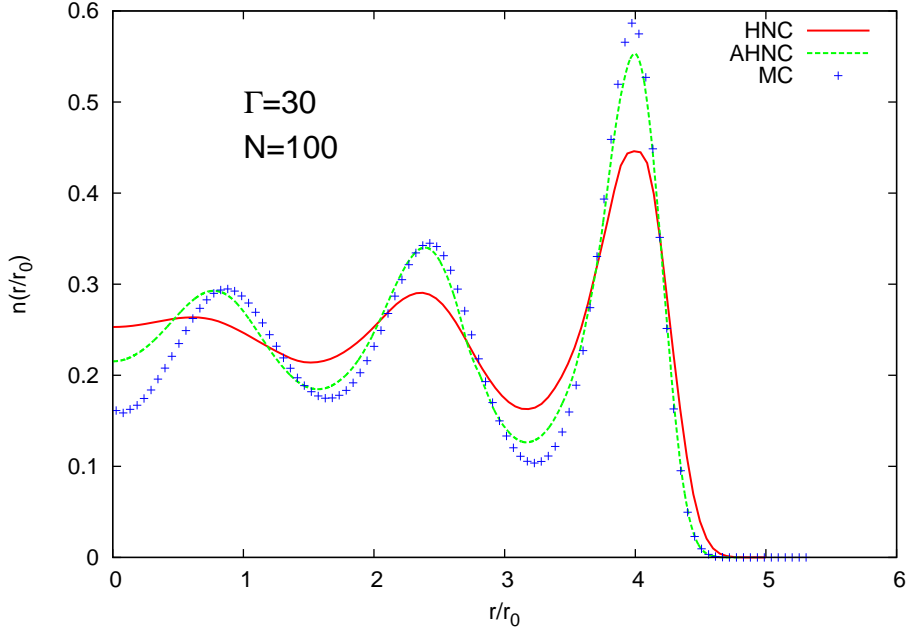}
\caption{Comparison of HNC and adjusted HNC (AHNC) to Monte Carlo (MC)
results. Conditions are $\Gamma=30$ and $N=100$.}
\label{fig1}
\end{figure}

The HNC density profile is very accurate at weak to moderate coupling, $%
\Gamma <10$, but has significant quantitative errors as the shell structure
develops at larger values of the coupling. This is illustrated in Figure \ref%
{fig1} for $\Gamma =30,N=100$. It was proposed in reference \cite{Wrighton09}
to improve the HNC by a phenomenological representation of the bridge
functions in the form%
\begin{equation}
B(r|n)\rightarrow \lambda (\Gamma )\beta V_{0}(r),  \label{5}
\end{equation}%
where $V_{0}$ is the harmonic potential for $B_{\mathrm{T}}$ or
Coulomb potential for $B_{\mathrm{OCP}}$. Also $\lambda (\Gamma )$
is chosen to interpolate between $\lambda (0)=0$ and some constant
$\lambda (\infty )$. This approach was first introduced by Ng
\cite{Ng} for the OCP pair distribution. Subsequently, Rosenfeld and
Ashcroft have defined related "modified" HNC approximations using
the bridge function as a fitting function \cite{Rosenfeld78}. The
advantage of the form (\ref{5}) is that the effect of the bridge
functions is to simply "renormalize" the external potential in both
eqs. (\ref{3}) and (\ref{4}), so that the original HNC approximation
is regained except with an effective coupling constant $\Gamma
^{\prime }$ defined by

\begin{equation}
\Gamma ^{\prime }=\left[ 1+\lambda (\Gamma )\right] \Gamma   \label{6}
\end{equation}%
In the original work of Ng, he obtained agreement with the Monte Carlo data
for $g_{\mathrm{OCP}}$ to within a few percent using $\lambda (\infty )=0.6$%
. That same value for $\lambda (\infty )$ has been used in reference \cite%
{Wrighton09} and in the results presented here.

The improvement gained from this AHNC is also shown in Figure 1. All three
curves agree regarding the number of peaks and the location of the peaks.
Also, there is quantitative agreement regarding the number of charges in
each shell (not shown). However, the HNC and AHNC results give very
different heights and widths of the peaks themselves. Comparison to the MC
results makes it clear that the bridge function effects are important at
this value of the coupling constant. Similar improvement has been
illustrated for $\Gamma =20$ and $40$ for several values of $N$ \cite%
{Wrighton09}.

\begin{figure}[tbp]
\includegraphics{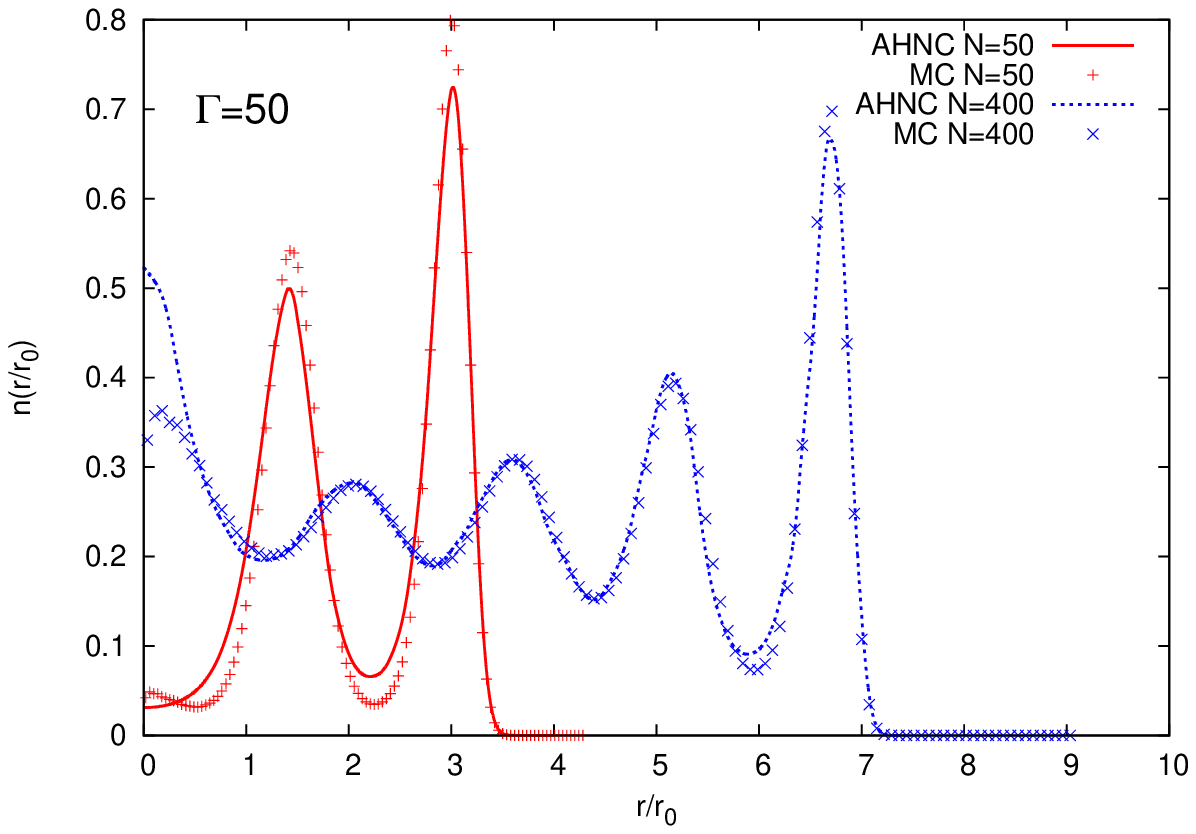}
\caption{Comparison of density profile for AHNC and MC, for $\Gamma =50$, $%
N=50$ and $400$ }
\label{fig2}
\end{figure}

\begin{figure}[tbp]
\includegraphics{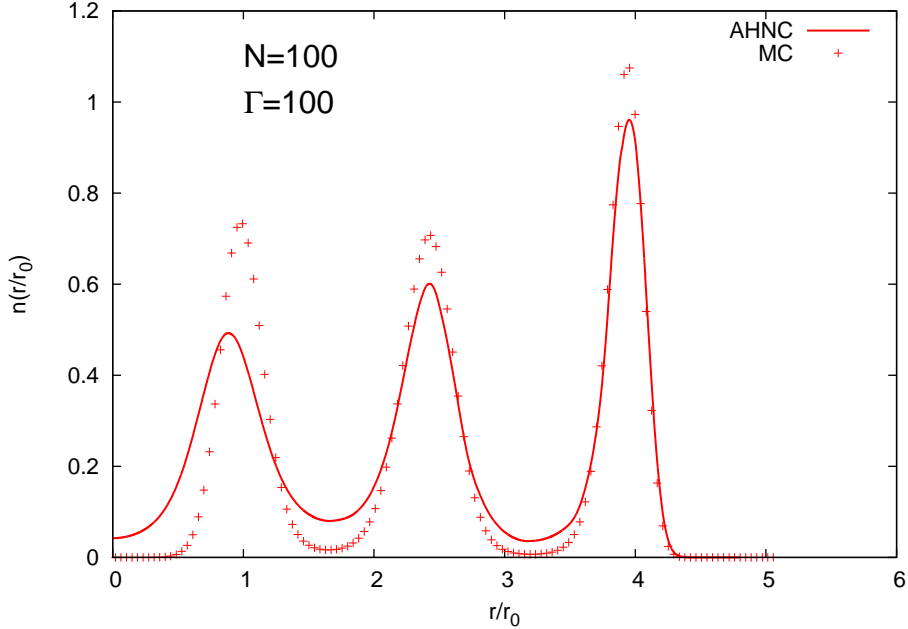}
\caption{Comparison of density profile for AHNC and MC results for $%
\Gamma=100$ and $N=100$.}
\label{fig3}
\end{figure}

It remains to explore how well this approach extends to still larger values
of $\Gamma $, motivated by its possible utility to study freezing into a
spherical crystal. \ As a first step in that direction results are presented
and discussed here for $\Gamma =50,100$. Figure ({\ref{fig2}) shows the
density profile at $\Gamma =50$ for two quite different particle numbers $%
N=50$ and $N=400$. For small enough particle number, $N\lesssim 20$,
there is only one shell of finite radius near the point at which the
density vanishes \cite{Ludwig2005}. The latter is fixed by the
condition that the Coulomb force on a test charge balances that of
the trap, i.e. $r_{max}=N^{1/3}$. A second shell is present at
$N=50$, while three additional shells are
observed for $N=400$ (as well as a population at the origin \cite{bonitz2006}%
). The agreement between MC and AHNC is seen to be quite good in both cases
except at very small }$r$ for which the both HNC and AHNC fail even at
moderate coupling.

\begin{figure}[tbp]
\includegraphics{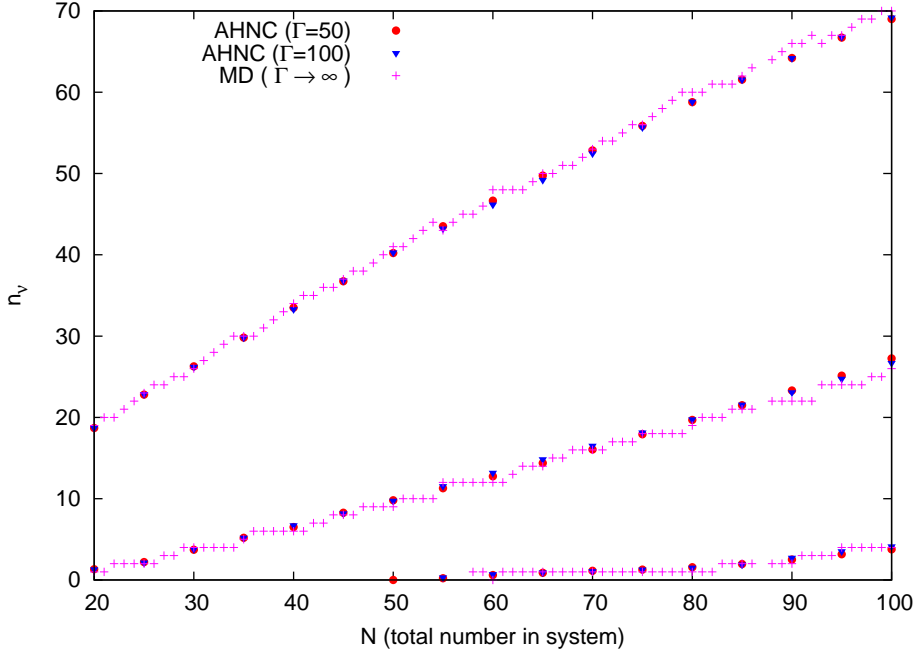}
\caption{Comparison of shell populations from AHNC and MD as a
function of particle number $N$.} \label{fig4}
\end{figure}

\begin{figure}[tbp]
\includegraphics{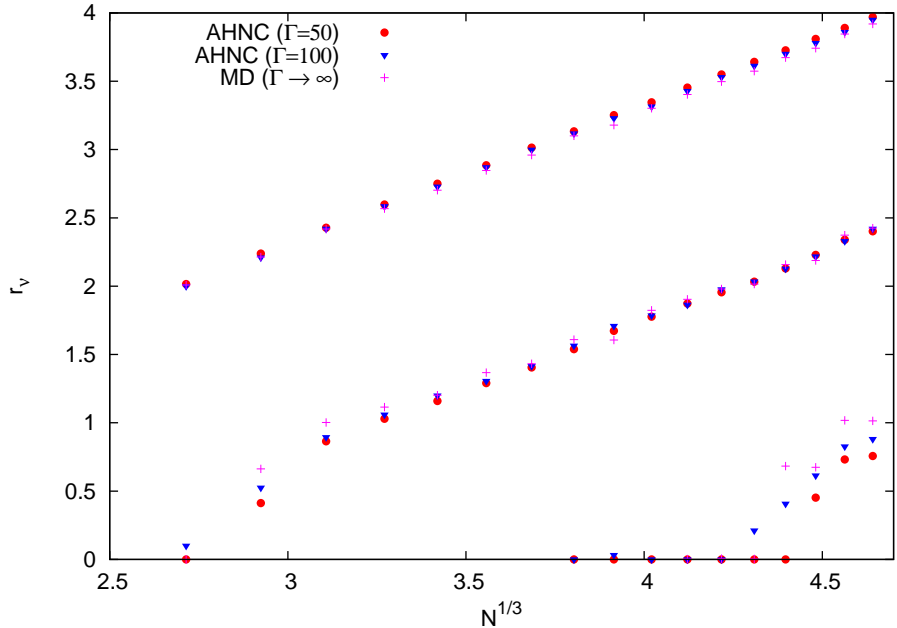}
\caption{Comparison of shell radii from AHNC and MD as a function of
particle number $N$.} \label{fig5}
\end{figure}

The discrepancies in amplitudes increase somewhat at very strong coupling.
This is illustrated in Figure (\ref{fig3}) showing the comparison of MC and
AHNC results for $\Gamma =100,N=100$. The error in peak height grows from
about $10\%$ for \ the outer shell to about $30\%$ for the inner shell.
Still the overall agreement remains quite good. To explore the conditions
for freezing into spherical crystals, even stronger coupling conditions, $%
\Gamma \sim 200$ are expected to be relevant. However, several
important features of the crystal ground state are already evident
from the fluid phase AHNC results at the coupling conditions
studied. Figure (\ref{fig4}) shows a comparison of shell populations
from ground state annealed MD simulations \cite{Arp} with those from
AHNC. The latter have been computed for $20\leq \Gamma \leq 100$,
showing a very weak dependence on coupling strength. This is
somewhat unexpected since the amplitudes and widths of the shells
are quite sensitive to $\Gamma $. The shell populations and
appearance of shells is clearly quite similar in the fluid and
crystal phases. Figure (\ref{fig5}) shows a corresponding comparison
of shell locations as a function of $N$. Again the results are
remarkably close for both fluid and crystal phases.

The results for the ground state shown on Figures (\ref{fig4}) and (\ref%
{fig5}) can be understood theoretically on the basis of a shell model. This
represents the density as a sum of delta functions for particle positions on
shells of radii $r_{\nu }$ and occupancy $n_{\nu }$
\begin{equation}
n(\mathbf{r})=\sum_{\nu }n_{\nu }\frac{1}{4\pi r_{\nu }^{2}}\delta \left( r%
\mathbf{-}r_{\nu }\right) ,\hspace{0.25in}\sum_{\nu }n_{\nu }=N.  \label{7}
\end{equation}%
The values of $n_{\nu },r_{\nu }$ are determined by minimizing the total
energy \cite{Baumgartner,Kraeft06}. A very accurate representation of the
energy for this purpose is constructed on the basis of solutions to the
Thomson problem \cite{Thomson} for ground state configurations of Coulomb
charges on a single sphere \cite{Cioslowski08}. The intrashell energy is
chosen as the known Thomson energy for each $n_{\nu },r_{\nu }$, while the
intershell energy is taken as the monopole interactions for charges at the
corresponding Thomson positions for each shell. In this way the crystal
energy, shell populations, and locations are given very accurately. This
opens the possibility to test free energies associated with (\ref{3}) and (%
\ref{4}) for broken rotational invariance using parameterized densities
similar to the shell model (e.g., Gaussians centered at the Thomson
positions) to identify a freezing transition.

This work is supported by the Deutsche Forschungsgemeinschaft via SFB-TR 24,
and by the NSF/DOE Partnership in Basic Plasma Science and Engineering under
the Department of Energy award DE-FG02-07ER54946.

\end{document}